\documentclass[letterpaper,10pt]{article} 

\usepackage{opticameet3} 

\newcommand\authormark[1]{\textsuperscript{#1}}

\usepackage{amsmath,amssymb}
\usepackage[colorlinks=true,bookmarks=false,citecolor=blue,urlcolor=blue]{hyperref} 

\usepackage{amsmath,amssymb}
\usepackage[colorlinks=true,bookmarks=false,citecolor=blue,urlcolor=blue]{hyperref} 
\usepackage{opticameet3} 
\usepackage{hyperref}
\usepackage{xspace}
\usepackage{amssymb}
\usepackage{amsmath,amssymb,amsfonts}
\usepackage{statmath}
\usepackage[figurename=Fig.]{caption}
\usepackage{algorithmic}

\hypersetup{
    colorlinks = false,
    hidelinks}

\begin{document}

\title{LuxNAS: A Coherent Photonic Neural Network Powered by Neural Architecture Search}

\copyrightyear{2025}
\vspace{-0.25 in}
\author{Amin Shafiee\authormark{*}, Febin Sunny\authormark{*}, Sudeep Pasricha, and Mahdi Nikdast}

\address{Department of Electrical and Computer Engineering, Colorado State University, Fort Collins, CO, USA\\}
\vspace{-0.05 in}
\email{\authormark{*} Authors with equal contribution.} 

\vspace{-0.25 in}
\begin{abstract}
 We demonstrate a novel coherent photonic neural network using tunable phase-change-material-based couplers and neural architecture search. Compared to the MZI-based Clements network, our results indicate 85\% reduction in the network footprint while maintaining the accuracy. 
\end{abstract}
\vspace{-0.05 in}
\section{Introduction}
\begin{figure}[htbp]
\vspace{-0.1 in}
    \centering
    \includegraphics[width=1\textwidth]{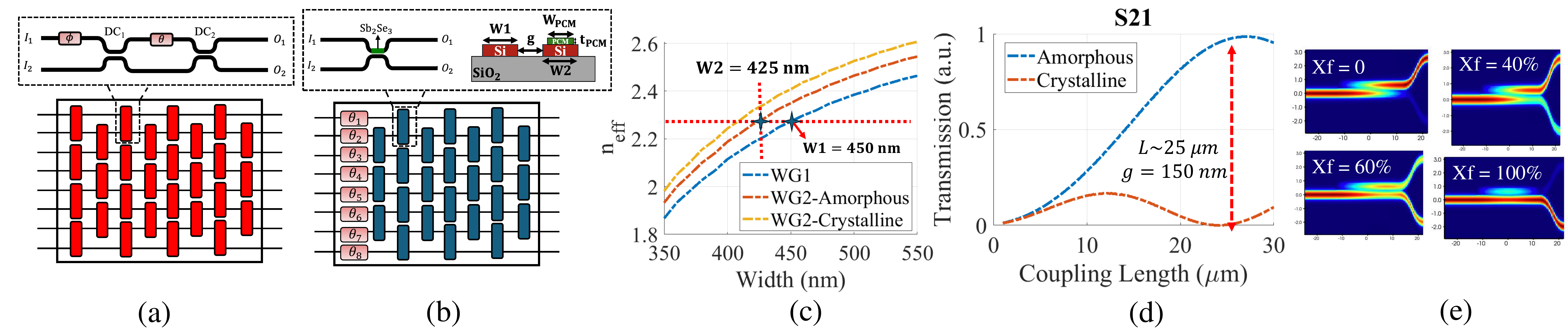}
    \vspace{-0.3 in}
    \caption{An 8$\times$8 (a) MZI-based Clements network and (b) LuxNAS network performing the same unitary transformation using Sb$_2$Se$_3$ loaded tunable DCs. (c) The effective index of a strip waveguide (WG1) and Sb$_2$Se$_3$-loaded waveguide (WG2). (d) Eigen-mode expansion simulation for S21 parameter as a function of the DC coupling length. (e) Field profile of the DC when Sb$_2$Se$_3$ is partially or fully crystallized and the coupling length is 25~$\mu$m. }
    \label{fig3_sin}
     \vspace{-0.35 in}
\end{figure}

\vspace{-0.05 in}
Silicon-photonic-based neural networks (SP-NNs) promise high energy efficiency and speed for performing computationally intensive multiply-and-accumulate (MAC) operations with high parallelism in the optical domain \cite{ghanaatian2023variation,cheng2020silicon}. Among various SP-NN implementations, coherent SP-NNs, which operate on a single wavelength, possess a distinct advantage over noncoherent SP-NNs due to higher computational density and performance. Yet, current coherent SP-NNs often require cascading a large number of active Mach--Zehnder Interferoemeters (MZIs), each with two integrated phase shifters (see Fig. \ref{fig3_sin}(a)). This limits the scalability of the network due to increased active power consumption and a large footprint. In this paper, we leverage the dynamic optical properties of the Sb$_2$Se$_3$ phase-change material (PCM) to first design a tunable directional coupler (DC) co-designed with an efficient phase shifter to selectively split and combine the optical inputs by tuning the crystalline fraction of the PCM \cite{xu2019low,zhang2024tunable}. We cascade several tunable PCM-based DCs with conventional phase shifters---all programmed based on a proposed neural architecture search (NAS)---to realize a multiplier network, called LuxNAS (see Fig. 1(b)), to perform MAC operations with zero active power consumption and significantly compact footprint.

\section{Coherent LuxNAS Network: Device Design and Neural Architecture Search}
\vspace{-0.05 in}
The schematic design of LuxNAS based on tunable PCM-based DCs is shown in Fig. \ref{fig3_sin}(b). By adjusting the crystallization fraction of the Sb$_2$Se$_3$ cells co-integrated with DCs, different degrees of interference can be achieved in the output of the network. The effective refractive index for a silicon-on-insulator (SOI) strip waveguide and a Sb$_2$Se$_3$ loaded strip waveguide (see Fig. 1(b)) when the PCM has amorphous and crystalline state is depicted in Fig. \ref{fig3_sin}(c). Note that Sb$_2$Se$_3$ is 30-nm thick and its width is always 100~nm shorter than the underlying waveguide. Observe that for the PCM-loaded waveguide (WG2), when the  waveguide width is 425~nm and the PCM is in the amorphous state, the effective index is almost the same compared to the width of 450~nm for the strip waveguide (WG1). Therefore, the phase match condition can be met. Note that the phase match condition does not hold as the PCM's state changes. Lumerical Eigen-Mode Expansion (EME) simulations were used to optimize the coupling of the light into the PCM-loaded waveguide when the PCM is in the amorphous state. S21 parameter as a function of the DC's length is depicted in Fig. \ref{fig3_sin}(d). Considering a 150-nm gap between the waveguides, 25~$\mu$m of the coupling length is required to fully couple the light into the amorphous PCM-loaded waveguide. The electric field profiles of the Sb$_2$Se$_3$-loaded tunable DC for different crystallization fractions are depicted in Fig. \ref{fig3_sin}(e).
\begin{figure}[htbp]
\vspace{-0.1 in}
    \centering
    \includegraphics[width=1\textwidth]{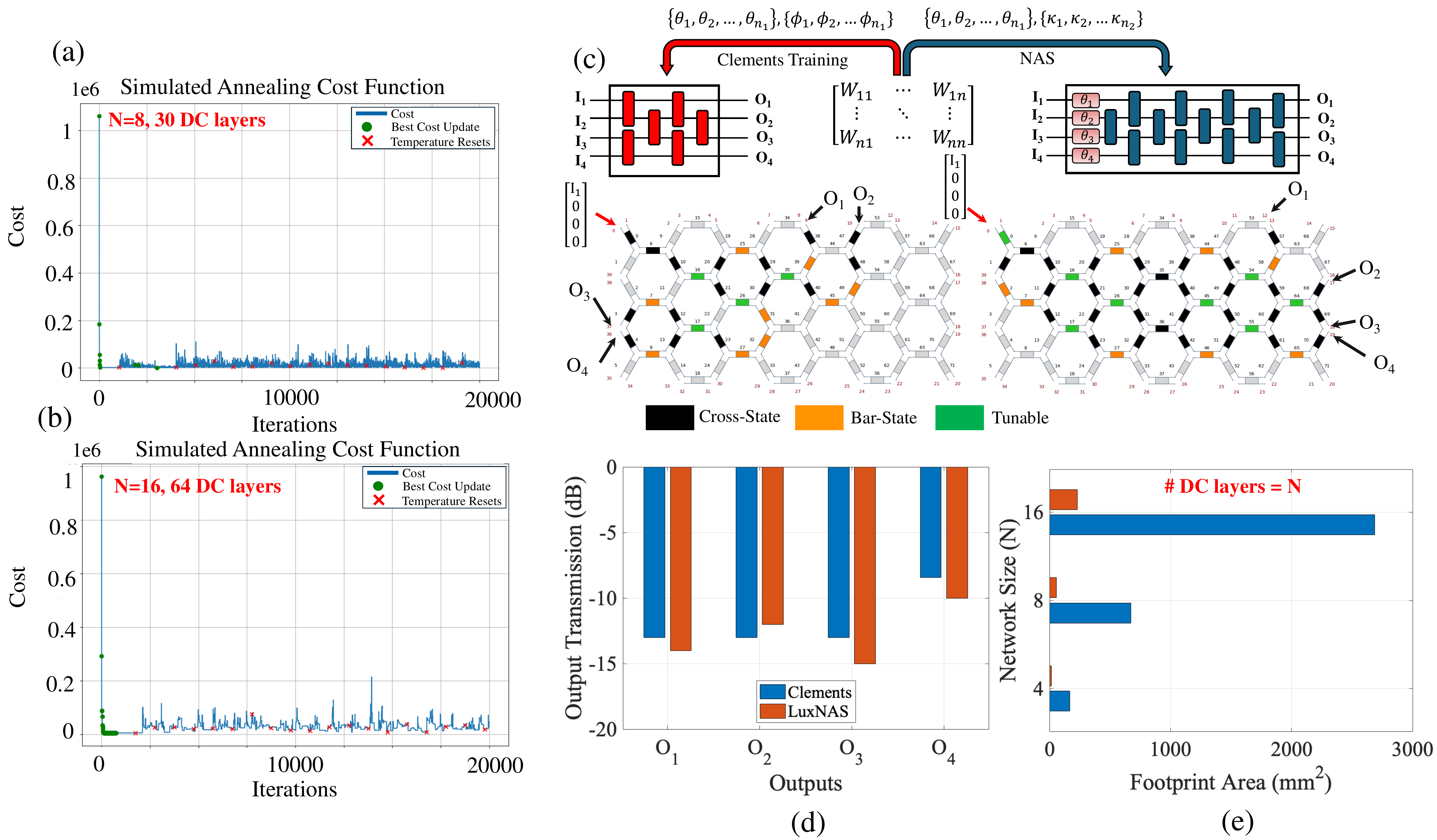}
    \vspace{-0.3 in}
    \caption{ Cost function versus the iterations for an (a) 8$\times$8 and a (b) 16$\times$16 LuxNAS network. (c) Emulated 4$\times$4 LuxNAS and Clements networks using an iPronics SmartLight processor. $n_1$ and $n_2$ denote the total number of MZIs and DCs in the Clements and LuxNAS, respectively. (d) Output optical transmission of the two emulated networks on the iPronics platform. (e) Footprint comparison between MZI-based Clements and LuxNAS networks.}
    \label{fig2}
     \vspace{-0.35in}
\end{figure}
 
 For NAS, a hybrid optimization combining Simulated Annealing (SA) for global exploration and Limited-memory Broyden–Fletcher–Goldfarb–Shanno with Box constraints (L-BFGS-B) is employed for local refinement, to optimize parameters of photonic mesh designs (coupling coefficients of the DCs ($\kappa$) and the phase shifters ($\theta_i$); see Fig. 1(b)). The hybrid strategy ensures faster convergence and improved computational efficiency. 
 The optimization minimizes a cost function that encapsulates fidelity, unitarity \cite{amin_jlt}, and smooth parameter transitions, facilitating robust photonic mesh design. The mesh structure is parameterized using coupling coefficients ($\kappa(x_f)$) and phase shifts ($\theta$), which are iteratively tuned to approximate a target transfer matrix $T'$ to the expected transfer matrix $T$.
The cost function, $C(\kappa, \theta)$, combines multiple terms to achieve optimization goals: $C(\kappa, \theta) = \alpha \cdot \left(\frac{1}{F(T,T')}\right)^2 + \beta \cdot \|\Delta \kappa\|^2 + \beta \cdot \|\Delta \theta\|^2 + \gamma \cdot \|T' \cdot T^{\dagger} - I\|^2$. Here, $F(T, T')$ represents the fidelity between the target matrix $T$ and the current matrix $T'$, $\Delta \kappa$ and $\Delta \theta$ are parameter changes from previous iterations, $\alpha$, $\beta$, and $\gamma$ are weight coefficient, and the Frobenius norm $\|T' \cdot T^{\dagger} - I\|^2$ penalizes deviations from unitarity ($T^{\dagger}$ being the Hermitian conjugate of $T'$). We performed an initial exploration using the baseline model of LuxNAS and Stochastic Gradient Decent (SGD) to estimate good weight coefficients ($\alpha$, $\beta$, and $\gamma$) for the cost function.

\vspace{-0.1 in}
\section{Results and Discussions}
\vspace{-0.05 in}

The cost function of the hybrid optimization as a function of number of iterations for two examples of an 8$\times$8 network with 30 tunable DC layers and a 16$\times$16 network with 64 tunable DC layers is shown in Figs. \ref{fig2}(a) and 2(b), highlighting the performance of LuxNAS for large networks. Note that the final fidelity was 0.75 and the relative variation distance (RVD) was less than 1. The inferencing accuracy of the SP-NN based on the optimized transformation matrix with respect to the equivalent trained Clements network on a linearly separable Guassian dataset \cite{amin_jlt} with the same size  was recalculated and the results showed less than 5\% accuracy drop. To validate the results, we emulated an example 4$\times$4 LuxNAS network with 8 tunable DC layers using an iPronics SmartLight Processor \cite{Ipronics}, by deploying the optimized $\kappa$ and $\theta$ values and then capturing the output transmission of the network. The emulated network could perform a 4$\times$4 transformation from its inputs to its outputs. The same procedure was performed for a trained 4$\times$4 Clements network on a linearly separable Gaussian dataset \cite{ghanaatian2023variation} using its trained phase values. Note that the trained weight matrix on Clements network was used for our optimization for LuxNAS (see Fig. \ref{fig2}(c)). Observe from Fig. \ref{fig2}(d) that when the input I$_1$ is excited (considered as an example), the normalized output optical transmissions (w.r.t. the input laser power at $-$6~dBm at 1.55~$\mu$m) for the two networks are very close to each other, which verifies the operation of LuxNAS. The footprint for LuxNAS and Clements networks when the $\#$DC layers$=$N (N is the number of network's input ports) is depicted in Fig. \ref{fig2}(e). Observe that LuxNAS can support the same MAC operation but with a significantly more compact (up to 85\% less) footprint. 
\vspace{-0.1 in}
\bibliographystyle{IEEEtran}
\bibliography{IEEEabrv,sample}

\end{document}